\documentclass[12pt]{article}
\usepackage[cp866]{inputenc}
\usepackage[russian]{babel}
\usepackage{amssymb}
\usepackage{amsmath}
\usepackage{epsfig}
\usepackage{multirow}
\usepackage[labelsep=period]{caption}

\newcommand{\be}{\begin{equation}}
\newcommand{\ee}{\end{equation}}
\newcommand{\bea}{\begin{eqnarray}}
\newcommand{\eea}{\end{eqnarray}}
\makeatother

\renewcommand{\refname}{}

\makeatletter
\renewcommand{\@biblabel}[1]{#1.}
\makeatother
\date{}

\begin{document}

\begin{center}
{\large\bf
ВЛИЯНИЕ ВСТРЯСКИ НА СКОРОСТЬ БЕЗНЕЙТРИННОГО ДВОЙНОГО 
ЭЛЕКТРОННОГО ЗАХВАТА В $^{164}$Er}  
\end{center}
\large
\bigskip
\begin{center}
Ф. Ф. Карпешин\footnote{ВНИИМ имени Д.И. Менделеева, Санкт-Петербург, Россия }$^{,\ast}$,  \frame{М. Б. Тржасковская}\footnote{НИЦ ``Курчатовский институт'' --- ПИЯФ, Санкт-Петербург,
Россия}
\end{center}  

 \vspace{0.1cm}
\begin{center}
$^\ast$E-mail: fkarpeshin@gmail.com
\end{center}

\smallskip \smallskip \bigskip   \noindent
\begin{abstract} Традиционно двойной безнейтринный
электронный захват рассматривают как резонансный процесс. Нами выполнены
расчеты вероятности встряски с ионизацией электронной оболочки,
имеющей место в случае  $^{164}$Er. Ее учет снимает требование резонанса, приводя к увеличению скорости захвата. Вклад нового механизма увеличивает скорость захвата в 5.6 раза по сравнению с  традиционным резонансно-флюоресцентым механизмом. Его учет  также повышает вероятность захвата электронов из более высоких оболочек, что необходимо учитывать в экспериментальном исследовании. Более того, учет встряски способен потенциально расширить список ядер-кандидатов на проведение экспериментов.
\end{abstract}
 \large
\bigskip

\begin{center}
ВВЕДЕНИЕ \\
\end{center}

Обнаружение скрытой материи во Веленной	 стимулирует развитие теорий за пределами стандартной модели. Как правило, они включают нарушение лептонного квантового числа, если не вводится специальных ограничений. Этим привлекается  большой интерес к исследованию двойных бета-процессов, включая 2$e$-распад ядра и захват им двух орбитальных электронов \cite{obzor}. В рамках стандартной модели лептонное квантовое  число сохраняется. Это исключает двойной безнейтринный бета-распад или $e$-захват. Последние становятся возможны только при наличии у нейтрино массы и при том, что нейтрино являются частицами майорановской природы. 
Однако обнаружение массы у нейтрино и их осцилляций уже ознаменовало наблюдение процессов за пределами стандартной модели. Таким образом, поиск безнейтринных двойных процессов должен дать ответ на вопрос о майорановской природе нейтрино. Из двух безнейтринных процессов большей скоростью распада  обладает  $2e\ 0\nu$-распад.  $2e\ 0\nu$-захват, хотя и уступает $2e\ 0\nu$-распаду несколько порядков по вероятности, более удобен с точки зрения детектирования. 

     Существенный момент состоит в том, что безнейтринный $2e$-захват  традиционно рассматривался как резонансный, поскольку ни одной частицы не испускается в результате ядерного превращения \cite{Wyceh}. В то же время закон сохранения требует передачи части освобождённой энергии третьему телу. В качестве такового выступает электронная оболочка атома. Закон сохранения энергии восстанавливается, например, вследствие излучения кванта флюоресценции, энергия которого включает в себя избыточную величину $Q$. Поэтому энергия этих квантов отличается от обычных квантов флюоресценции, которая имеет место как правило в ионизованных атомах с одной-единственной дыркой во внутренней оболочке. Обнаружение таких сателлитов в спектре  флюоресценции и может служить индикатором безнейтринного или двухнейтринного двойного электронного захвата \cite{h2,FK}. Таким образом,  амплитуда безнейтринного захвата перманентно включает в себя, наряду с собственно амплитудой 
$2e$-захвата,  радиационную вершину, что удлиняет период процесса на два порядка. Поэтому главный критерий прикован к изучению ядер с малым значением $Q$. В работе \cite{elis} был отобран список из трех наиболее подходящих кандидатов на экспериментальное исследование: $^{152}$Gd, $^{164}$Er и $^{180}$W. В данной работе мы уточняем вопрос о вероятности  процесса $^{164}$Er  $\to$ $^{164}$Dy, основываясь на новом механизме встряски, предложенном в работе \cite{nrl}.  

      Не требуя резонанса, механизм встряски может быть реализован независимо от величины $\Delta$, поскольку его вклад медленней убывает с увеличением дефекта резонанса, чем обычного резонансно-флуоресцентного механизма.  Восстановление закона сохранения   энергии происходит вследствие ионизации электронной оболочки. При этом избыток энергии уносится вылетевшим электроном. Оценки эффективности встряски были выполнены в работе \cite{nrl} на примере распада 
$^{152}$Gd $\to$ $^{152}$Sm, обладающего минимальным дефектом резонанса среди известных кандидатов. Количественно вклад нового механизма оказался на уровне 23\% по сравнению с традиционным механизмом. Ожидается, однако, что вклад встряски увеличивается с ростом дефекта резонанса, так что он может стать доминирующим  в случае  захвата с большим энерговыделением. В данной работе мы рассматриваем как раз такой случай на примере процесса 
$^{164}$Er  $\to$ $^{164}$Dy, для которого  $\Delta$ = 6.82 кэВ. Результаты подтверждают ожидания. Учет встряски сокращает ожидаемый период процесса почти в шесть раз. В следующем разделе мы напомним основные формулы. Результаты расчетов   приводятся в разд. 3. Разд. 4 посвящен обсуждению результатов, полученных в настоящей работе. \\

\begin{center}
{\bf 1. Сравнение двух механизмов безнейтринного $2e$-захвата: физические принципы и формулы для расчета} \\
\end{center}

	При 2$e\ 0\nu$-захвате атом остается в целом нейтральным. Поэтому энерговыделение определяется разностью масс нейтральных атомов --- начального $M_1$ и дочернего $M_2$\footnote{Мы используем релятивистскую систему единиц $\hbar = c = m_e=1$, $m_e$ --- электронная масса, если не указано иначе.}:
\be
Q = M_1 - M_2 \,.	\label{Q}
\ee
Однако процесс с полным энерговыделением (\ref{Q}) мог бы реализоваться лишь при захвате самого внешнего, валентного электрона. Гораздо более вероятен захват внутренних электронов, плотность которых на ядре выше, соответственно атом остается в возбуждённом состоянии с энергией $E_A^\star$ и двумя дырками в раздутой оболочке \cite{FK}. Соответственно вместо (\ref{Q}) реализуется эффективное энерговыделение
\be
Q_\text{eff}= M_1 - M_2 -E_A^\star =Q -E_A^\star \,.   \label{defect}
\ee  
Процесс энергетически возможен при $Q>0$, но $Q_\text{eff}$ может быть и отрицательным: избыток энергии может как прибавляться, так и вычитаться из энергии кванта-сателлита. Именно $Q_\text{eff}$  выступает в роли дефекта резонанса $\Delta = |Q_\text{eff}|$. 

	Формулу для  резонансного механизма запишем в однополюсном приближении, воспользовавшись традиционной моделью. Формула  получается  (например,  \cite{shb}) умножением квадрата амплитуды  собственно захвата $\Gamma_{2e}$, который играет роль образования входного состояния,  на резонансный фактор  Брейта---Вигнера
\be
\Gamma_{2e}^{(\gamma)} = \Gamma_{2e} B_W \,,  \label{Wr}
\ee
где
\be
B_W = \frac{\Gamma/2\pi}{\Delta^2+(\Gamma/2)^2}\,.   \label{BW}
\ee
В (\ref{BW}) $\Gamma$ ---  ширина раздутого  состояния дочернего атома с двумя электронными дырками, равная сумме ширин каждой из дырок. 

\begin{figure}[!bt]
\centerline{ \epsfxsize=10cm\epsfbox{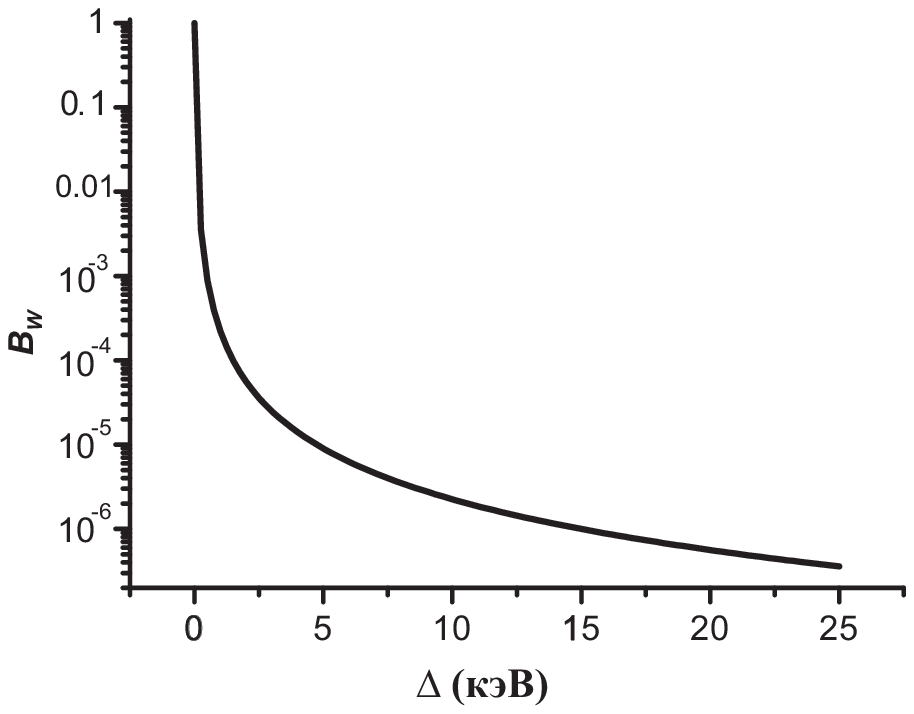}}
\caption{\footnotesize {Типичная зависимость резонансного фактора 
Брейта---Вигнера (\ref{BW}) от дефекта резонанса $\Delta$. }} \label{Deltaf}
\end{figure}

	Типичное значение $\Gamma \approx$ 30 эВ. Для иллюстрации на рис. \ref{Deltaf} показан масштаб изменения фактора $B_W$ в зависимости от 
дефекта резонанса. Он спадает в два раза при $\Delta$ = 1.5 кэВ. Только один кандидат известен с подходящим значением: $^{152}$Gd, у которого $\Delta$ = 0.91 кэВ. В остальных известных случаях $\Delta$ составляет величину по крайней мере от нескольких кэВ до одного-двух десятков кэВ, при этом фактор 
Брейта---Вигнера спадает до шести порядков величины. 

     Встряска энергетически возможна только при положительных $Q_\text{eff}>0$ с оболочек, чей потенциал ионизации в дочерненм атоме, с двумя вакансиями в электронной оболочке, $I_i$ удовлетворяет условию 
\be
I_i<Q_\text{eff}\,, 
\ee	а энергия электронов встряски определится их разностью
\be
E_\text{sh} = Q_\text{eff}-I_i\,. 
\ee	

	Встряска возникает вследствие изменения внутриатомного электростатического потенциала $V_{Z}(r)$ в исходном атоме на потенциал в дочернем атоме $V_{Z-2}(r)$. Поэтому одноэлектронные волновые функции начального и конечного атомов неортогональны, даже  с разными квантовыми числами. Для расчета вероятности встряски наиболее важна неортогональность волновых функций: $\phi_i(r)$ в родительском атоме и $\psi_f(r)$ электрона встряски с энергией $E_\text{sh}$ в дочернем атоме. Обозначим изменение потенциала  $\Delta V(r)\equiv V_{Z}(r)-V_{Z-2}(r)$.  
Таким образом, волновая функция $\phi_i(r)$ принадлежит родительскому нейтральному атому в основном состоянии, а $\psi_f(r)$ вычисляется в поле дочернего атома с тремя вакансиями. 
Тогда амплитуда встряски будет  \cite{land}
\be
F_{\text{sh}}(\Delta) = \langle \psi_f|\phi _i\rangle  \,.   \label{sh1}
\ee
 Полную  амплитуду можно представить, аналогично (\ref{Wr}), в виде произведения
\be
F_{2e}^{\text{ (sh)}} = F_{2e}F_{\text{sh}}(\Delta)  \,.	\label{W2bs}
\ee   
Сравнивая (\ref{W2bs})  с  (\ref{Wr}), получим относительную поправку к вероятности распада в единицу времени:
\be
G = \Gamma_{2e}^{\text{(sh)}} / \Gamma_{2e}^{(\gamma)} =
\sum_i N_i |\langle \psi_f|\phi_i\rangle  |^2 / B_W \equiv  \sum_i N_i |F_{\text{sh}}(\Delta)|^2 /B_W \,.  \label{rtio}
\ee

      В рамках резонансного механизма основной вклад возникает от захвата двух $L_1$-электронов в атомах $^{164}$Er, в результате чего они   превращаются в атомы $^{164}$Dy. Захват более низких электронов энергетически невозможен, а более высоких подавляется как уменьшением их плотности на ядре, так и падением фактора Брейта---Вигнера вследствие увеличения $Q_\text{eff}$. В нерезонансном механизме встряски уменьшение электронной плотности на ядре например, при захвате из $M_1$-оболочки наоборот, частично компенсируется увеличением $Q_\text{eff}$, так как открывается канал встряски электронов с $L_1$-оболочки. Это приводит к тому, что вероятность захвата из более высоких оболочек с учетом встряски, как мы увидим, оказывается даже выше, чем вероятность традиционного резонансного механизма. В случае, если происходит $ik$-захват с более высоких оболочек $i$, $k$, то  можно рассчитать фактор ускорения по сравнению с наиболее вероятным резонансным $L_1L_1$-захватом по формуле 
\be
G_{ik} = \frac{\rho_i(0)\rho_k(0)}{\rho_{L_1}^{\ 2}(0)}
\sum_j N_j |F_{\text{sh}}^{(j)}(|Q_\text{eff}^{(ik)}|)|^2 /B_W \,.  \label{rth}
\ee
Здесь суммирование по-прежнему проводится по всем оболочкам $j$, энергетически разрешенным для встряски,  с числами заполнения $N_j$.
$F_{\text{sh}}^{(j)}(|Q_\text{eff}^{(ik)}|)$ --- по-прежнему интеграл перекрытия волновых функций электрона на оболочке $j$ и электрона в континууме, но вычисленный при фактическом энерговыделении $Q_\text{eff}^{(ik)}$, соответствующем $ik$-захвату. В случае наиболее вероятного $L_1L_1$-захвата самой нижней оболочкой, с которой идет встряска, является $M$-оболочка. Если захват одного из электронов происходит с $M$-оболочки, то величина $Q_\text{eff}^{(LM)}$ увеличивается на  разность потенциалов ионизации $L$- и $M$-оболочек. Это автоматически открывает канал встряски с $L$-оболочки ($L_1$, $L_2$, $L_3$), что приводит к скачкообразному увеличению вероятности встряски. 

\begin{center} {\bf 3. Результаты расчетов} \\
\end{center}

    Расчеты по формулам (\ref{sh1}), (\ref{rtio}) выполнены в одноэлектронном приближении с помощью комплекса программ  RAINE \cite{RAINE,AD}. Волновые функции электронов и их энергии вычислялись самосогласованным  методом Дирака---Фока. С целью лучшего понимания физики процесса были рассчитаны матричные элементы (\ref{sh1}) для ряда гипотетических значений  $\Delta$ от 0.05  до 20 кэВ  для всех электронов, чьи потенциалы ионизации меньше заданной величины $\Delta$ и которые следовательно вносят вклад в амплитуду нерезонансного механизма нашего процесса.

            Результаты расчетов представлены на рис. 2 -- 5, а также в таблицах 1, 2.
Наши волновые функции нормированы на единицу для дискретных состояний и по шкале энергий --- в континууме.  Поэтому квадрат матричного элемента $F_{\text{sh}}(\Delta)$ имеет размерность, обратную энергии. Матричные элементы и факторы Брейта---Вигнера приведены ниже в релятивистской системе единиц. 

\begin{figure}[!bt]
\centerline{ \epsfxsize=10cm\epsfbox{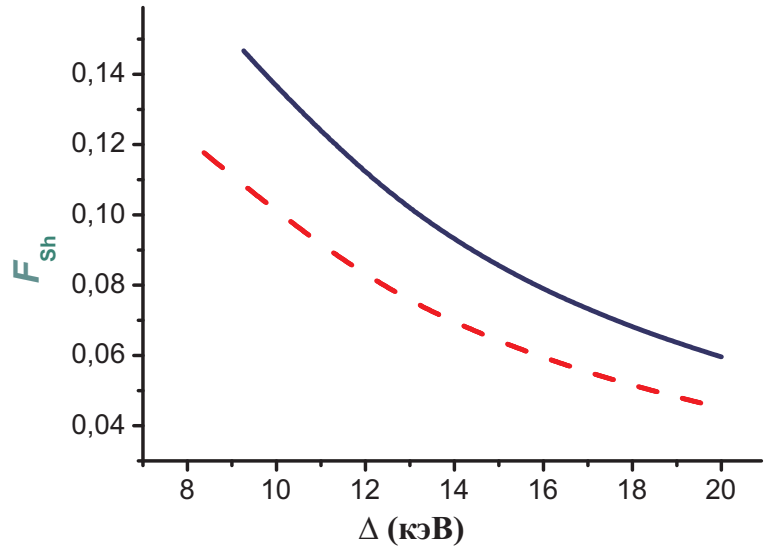}}
\caption{\footnotesize {Матричные элементы $F_{sh}$   для  $2p_{1/2}$- (сплошная кривая) и 
$2p_{3/2}$- (штриховая линия) подоболочек атома $^{164}$Dy в зависимости от дефекта резонанса $\Delta$.}} \label{2pf}
\end{figure}

\begin{figure}[!bt]
\centerline{ \epsfxsize=10cm\epsfbox{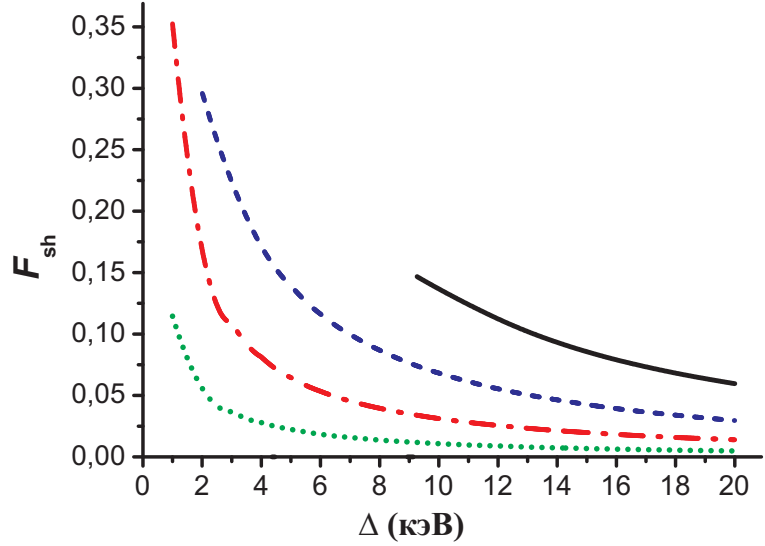}}
\caption{\footnotesize {Матричные элементы $F_{sh}$   для  
$np_{1/2}$-подоболочек атома $^{164}$Dy: $n$=2 (пунктир), $n$=3 (штрих-пунктир), $n$=4 (штриховая линия) и $n$=5 (сплошная кривая).}} \label{npf}
\end{figure}

\begin{figure}[!bt]
\centerline{ \epsfxsize=10cm\epsfbox{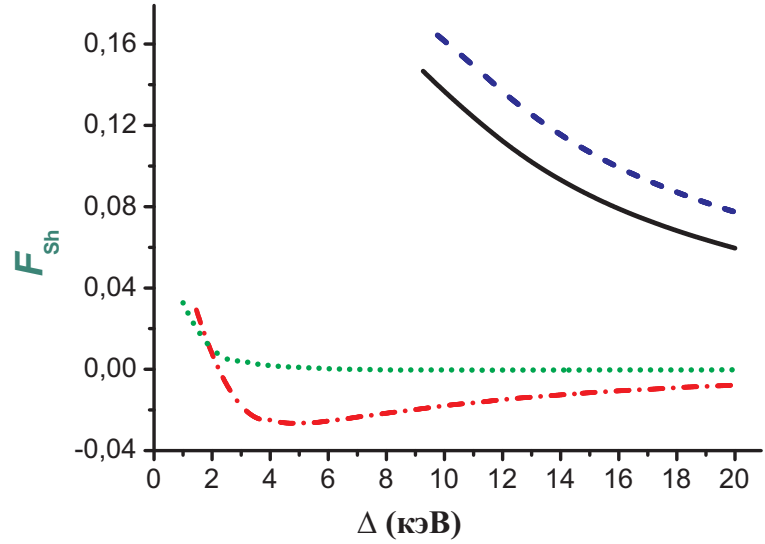}}
\caption{\footnotesize {Сравнение вкладов оболочек в амплитуду встряски в зависимости от орбитального углового момента. Матричные элементы $F_{sh}$   для  $2s$-подоболочки (штриховая линия), $2p_{1/2}$-подоболочки (сплошная кривая), $3d_{3/2}$ (штрих-пунктир) и 
$4f_{5/2}$ (пунктир) подоболочек атома $^{164}$Dy.}} \label{restf}
\end{figure}

Чем ближе оболочка к ядру, тем больший вклад дает она во встряску, если та  не запрещена энергетически.  Это иллюстрируется на рис. \ref{2pf}, на котором 
приведены матричные элементы $F_{sh}$   для  $L_2$- и 
$L_3$-подоболочек, представляющих компоненты тонкой структуры $2p$-орбитали. Кривые имеют различные пороги:  9.264   и 8.358   кэВ, соответственно. Оба порога выше эффективного энерговыделения, поэтому в данном случае ни одна кривая не вносит вклада во встряску в наиболее вероятном случае $L_1L_1$-захвата. На рис. \ref{npf} показаны матричные 
элементы для $2p_{1/2}$ --- $5p_{1/2}$-оболочек. Матричные элементы для остальных оболочек проиллюстрированы на рис. \ref{restf}.

\begin{figure}[!bt]
\centerline{ \epsfxsize=10cm\epsfbox{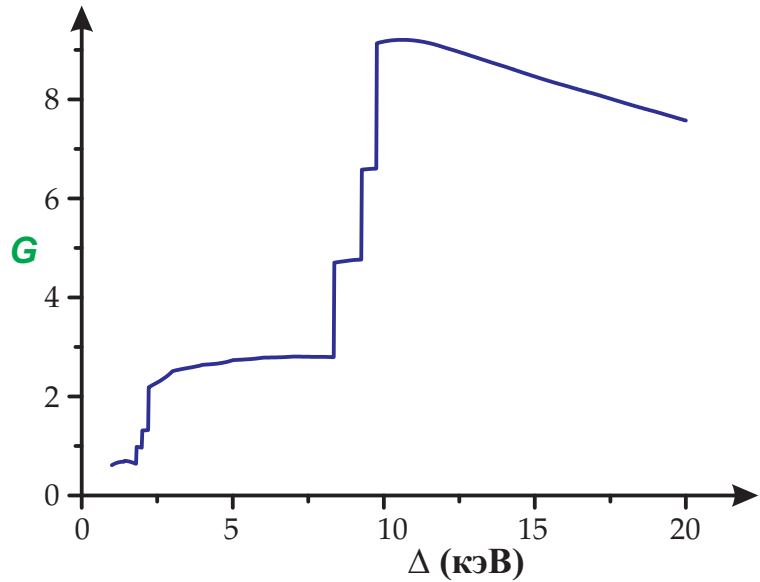}}
\caption{\footnotesize{Выигрыш встряски $G$ (\ref{rtio}) по сравнению с 
резонанcно-флюоресцентноым механизмом  в вероятности двойного
безнейтринного $L_1L_1$-захвата в $^{164}$Er в зависимости от дефекта
резонанса $\Delta$.}}
\label{gainf}
\end{figure}

     Суммарный фактор ускорения, соответствующий   механизму встряски  от всех электронов, относительно резонансного механизма  представлен на рис. \ref{gainf}. Вероятность этого процесса носит резко выраженный 
ступенчатый характер благодаря тому, что с увеличением $Q$ подключаются все  более глубокие оболочки, причем чем более глубоко лежит оболочка, тем большую величину составляет ее вклад на пороге. Как и ожидалось, основной вклад происходит от электронов $s$- и 
$p$-оболочек. 
Видно, что при  малых $Q$  резонансный механизм доминирует. При фактическом значении $Q$ = 6.82 кэВ вклад нерезонансного механизма в три  раза превышает традиционное значение.

	Приведенные на рисунках (\ref{2pf}) -- (\ref{restf}) значения можно использовать для оценки вклада  встряски в случаях захвата электронов и с других, более высоких оболочек. Используя полную ширину $L_{1}$-дырки в атоме Dy: $\Gamma$ = 4.3 эВ \cite{AD}, получим по формуле (\ref{rth}) факторы ускорения для захвата с $L$-, $M$- и $N$-оболочек. Они приведены в таблице \ref{shtab}. 
Из приведенных результатов следует, что учет более высоких оболочек приводит к росту скорости захвата от 2.8 (в случае $L_1L_1$-захвата) до 4.6 раза.
\begin{table}
\caption{\footnotesize Вычисленный вклад различных оболочек в ускорение $2e0\nu$-захвата $G$. $\rho(0)$ --- нормированное на $L_1L_1$-захват произведение плотностей в центре ядра двух электронов, захваченных ядром \cite{band}. }
\begin{center}
\begin{tabular}{||c||c|c|c||}
\hline  \hline
Оболочка & 	$\Delta$ (кэВ) &	$\rho(0)$ &	$G$ \\
\hline  			
$LL$  &  	6.82  &	1 &	2.81 \\
$LM$   &	14   &	0.218 &	1.22   \\
$MM$   &   	21   &	0.048  &	0.20  \\
$LN$   &  	15.6   &	0.051   &	0.29  \\
$MN$   &	22.6   &	0.011    &	0.05   \\
NN  &	24.3  &	0.003  &	0.01  \\
\hline
Итого:   &	---  &	---  &	4.58  \\
\hline  \hline
\end{tabular}
\end{center}
\label{shtab}
\end{table}

	Анализ приведённых в таблице результатов подтверждает весомый вклад более высоких оболочек в вероятность захвата по механизму встряски. Он оказывается в 1.8 раза выше, чем вклад 
резонансно-флюоресцентного механизма. А полный выигрыш составляет 5.6 раза. \\

\begin{center} {\bf 4. Обсуждение результатов} \end{center}

     1. Учет  процессов встряски в безнейтринном двойном ядерном $e$-захвате сдвигает теоретические оценки значительно ближе к экспериментальным возможностям. Его особенность состоит в том, что это нерезонансный механизм. Поэтому можно ожидать, что он окажется более вероятным при распаде в ядрах, характеризуемых значительным энерговыделением. В случае безнейтринного механизма большое энерговыделение означает и большой дефект резонанса, вследствие чего значительно  уменьшается вероятность распада по резонансному механизму. Рассмотрение нерезонансного механизма позволяет значительно уточнить оценку периода распада. Приведенные выше расчеты  подтверждают это предположение: учет нового механизма увеличивает оценку вероятности двойного захвата в 5.6 раза по сравнению с традиционным резонансно-флуоресцентным механизмом в случае $^{164}$Er.    Учитывая полученную в работе \cite{shb} оценку периода полураспада этого ядра относительно $2e0\nu$-захвата  $2\times 10^{30}$ лет в расчете на эффективную массу нейтрино $m_{\beta\beta}$ = 1 эВ, получим уточненную оценку полупериода $T_{1/2}^{0\nu} \approx 3.6\times 10^{29} \left |\frac{\text{1 эВ}}{m_{\beta\beta}}\right |^2$  лет. 
     В таблице \ref{t3} мы сводим ожидаемые результаты для периодов полураспада трех вышеуказанных кандидатов: $^{152}$Gd, $^{164}$Er
     и $^{180}$W.
\begin{table}
\caption{\footnotesize Периоды полураспада $^{152}$Gd, $^{164}$Er
и $^{180}$W путем двойного безнейтринного $e$-захвата, согласно резонансно-флюоресцентному механизму и встряске.} 
\begin{center}
\begin{tabular}{||c||c|c|c||}
\hline   \hline
Ядра & $^{152}$Gd$\to ^{152}$Sm & $^{164}$Er  $\to$ $^{164}$Dy & 
$^{180}$W  $\to$ $^{180}$Hf  \\
\hline
Канал распада & KL & LL & KK \\
$\Delta$ (кэВ) & 0.910 & 6.82 & 12.5 \\
Резонансный период (годы) & 10$^{27}$ & $2\times 10^{30}$ & 
$3\times 10^{28}$ \\
Нерезонансный период & $8\times 10^{26}$ & $3.6\times 10^{29}$ & 
$3\times 10^{27}$ \\
\hline    \hline
\end{tabular}
\end{center}
\label{t3}
\end{table}
     Принимая во внимание, что период распада  другого кандидата на измерение $0\nu2e$-захвата $^{152}$Gd остается на четыре порядка короче \cite{nrl}, можно заключить, что он остается более вероятным кандидатом на постановку эксперимента, чем $^{164}$Er. Оценки, аналогичные приведенным выше, показывают, что в других случаях более тяжелых ядер с эффективным энерговыделением $\gtrsim$ 10 кэВ, в том числе $^{180}$W,  выигрыш составляет уже полный порядок величины.
Тогда ожидаемое время жизни  $^{180}$W относительно $2e\ 0\nu$-захвата оказывается всего в четыре раза больше, чем  $^{152}$Gd. Это может сделать его более предпочтительным кандидатов,  учитывая, что распространенность изотопа $^{152}$Gd в природе составляет всего 0.2\%. 

     2. Учитывая, что  учет  встряски приводит к активизации роли высших оболочек, можно ожидать, что спектр флюоресценции имеет более богатую структуру, чем два сателлита, возникающие при захвате обоих электронов только с одной или двух определенных оболочек. В случае $^{164}$Er, в 20\% случаев захват осуществляется с более высоких орбит, чем $L_1$. Это приводит к сдвигу энергии основных сателлитов. Более того, возникают сателлиты квантов флюоресценции, отвечающих переходам в состояния $M$- и  $N$-оболочек. Это обстоятельство необходимо учитывать в эксперименте. Можно его использовать для более надежной идентификации процесса.

     Подводя итог, можно заключить, что нерезонансный механизм двойного безнейтринного $e$-захвата представляет важный пример, в котором интереснейший процесс встряски проявляется удивительно ярко. 

\bigskip
Один из авторов (ФФК) хотел бы выразить признательность Ю. Н. Новикову за инициирующие обсуждения.

\newpage
\renewcommand\refname{\centerline{\normalfont СПИСОК ЛИТЕРАТУРЫ}}

\large
\newpage
\begin{center}
EFFECT OF SHAKE-UP ON THE RATE OF A NEUTRINOLESS DOUBLE
ELECTRONIC CAPTURE IN $^{164}$Er

\medskip
  F. F. Karpeshin\footnote{D.I. Mendeleyev Institute for Metrology.}, \frame{M. B. Trzhaskovskaya}
\footnote{National Research Center ``Kurchatov Institute'' --- Petersburg Nuclear Physics Institute.}
\bigskip

Abstract
\end{center}

Traditionally double neutrinoless
electronic capture is considered as a resonance process. We have fulfilled
shake-off probability calculations, leading to ionization of the electron shell, 
in the case of $^{164}$Er. Allowance for the shake-off removes the requirement of resonance, leading to an increase of the capture rate. The contribution of the new mechanism increases the capture rate by a factor of 5.6 compared to the conventional resonance fluorescence mechanism. It  also increases the probability of electron capture from higher shells, which must be foreseen in an experimental study. Moreover, effect of the shake-off can potentially expand the list of candidate nuclei for experiments.
\end{document}